\documentclass[aps,prl,twocolumn,showpacs]{revtex4}

\usepackage{graphicx}
\begin{document}


\title{Pressure induced transition from a spin glass to an itinerant ferromagnet in half doped manganite {\it Ln}$_{0.5}$Ba$_{0.5}$MnO$_{3}$ ({\it Ln}=Sm and Nd) with quenched disorder}

\author{N. Takeshita$^{1}$, C. Terakura$^{1}$, D. Akahoshi$^{1}$, Y. Tokura$^{1,2}$, and H. Takagi$^{1,3,4}$}
\affiliation{$^{1}$ Correlated Electron Research Center (CERC), 
 National Institute of Advanced Industrial Science and Technology (AIST), 
 Tsukuba, Ibaraki 305-8062, Japan}
\affiliation{$^{2}$ Department of Applied Physics, University of Tokyo, 
 7-3-1 Hongo, Bunkyo-ku, Tokyo 113-8656, Japan}
\affiliation{$^{3}$ Department of Advanced Materials Science, 
 University of Tokyo, 5-1-5 Kashiwa-no-ha, Kashiwa 277-8581, Japan}
\affiliation{$^{4}$ CREST, Japan Science and Technology Corporation (JST)}

\date{\today}

\begin{abstract}
The effect of quenched disorder on the multiphase competition has been investigated by examining the pressure phase diagram of half doped manganite {\it Ln}$_{0.5}$Ba$_{0.5}$MnO$_{3}$ ({\it Ln} = Sm and Nd) with A-site disorders. Sm$_{0.5}$Ba$_{0.5}$MnO$_{3}$, a spin glass insulator at ambient pressure, switches to a ferromagnetic metal with increasing pressure, followed by a rapid increase of the ferromagnetic transition temperature {\it T}$_{C}$. The rapid increase of {\it T}$_{C}$ was confirmed also for Nd$_{0.5}$Ba$_{0.5}$MnO$_{3}$. These observations indicate that the unusual suppression of the multicritical phase boundary in the A-site disordered system, previously observed as a function of the averaged A-site ionic radius, is essentially controlled by the pressure and hence the band width. The effect of quenched disorder is therefore much enhanced with approaching the multicritical region.
\end{abstract}

\maketitle

One of the hallmarks of strongly correlated electron physics is the multiphase competition produced by a delicate interplay among charge, spin and orbital degrees of freedom~\cite{ref1,ref2}. An appealing example of such multicriticality can be seen in perovskite manganites. Competing interactions/orders inherent in the manganites, such as double-exchange ferromagnetism vs. super-exchange antiferromagnetism and charge-orbital order vs. metallic state, tend to produce the multicritical state where external stimuli occasionally cause the dramatic phase conversion, e.g., between a metal and an insulator, or between a ferromagnet and an antiferromagnet~\cite{ref3}. The colossal magnetoresistance (CMR) phenomena~\cite{ref4} are believed to be a consequence of such phase competitions.\par
Recently, the vital role of quenched disorder in the critical region of manganites has been attracting considerable interest. The quenched disorder~\cite{ref5} may result in a phase separation~\cite{ref6} into the competing two ordered phases on various time-scales and length-scales, which substantially modifies the criticality and hence the response to the external field. The first suggestion came from the observation of relaxor-like behavior induced by Cr-doping on the Mn sites in the charge/orbital ordered (CO/OO) state~\cite{ref7,ref8}. This is highly likely associated with the ferromagnetic (FM) metallic clusters coexisting with the background CO/OO state. The effect of quenched disorder on the A-site was subsequently examined, which shows a drastic modification of the phase diagram as a function of the average A-site ionic radius.\par

\begin{figure*}[t]
\includegraphics[width=0.8\linewidth]{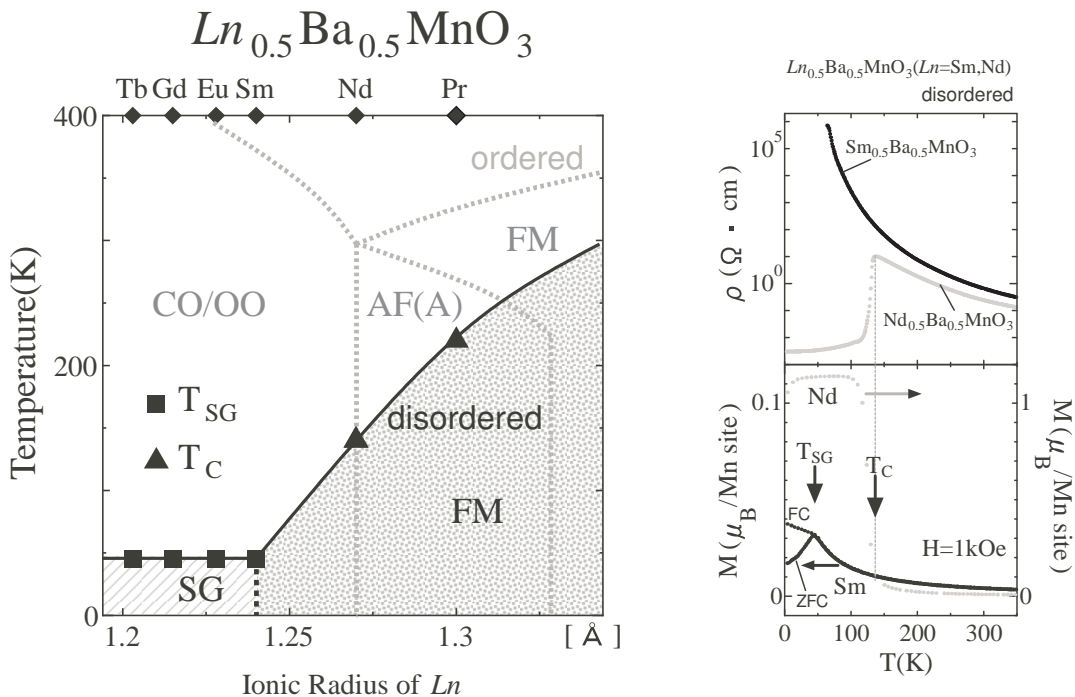}
\caption{Phase diagrams as a function of ionic radius of {\it Ln} ions are shown in the left panel for A-site ordered and disordered {\it Ln}$_{0.5}$Ba$_{0.5}$MnO$_{3}$. (Data taken from ref.~\cite{ref11} by Akahoshi {\it et al}.)
 The solid and dotted lines indicate those for disordered and ordered systems, respectively. FM, AF, SG, CO/OO denote ferromagnetic metal, antiferromagnet, spin glass and charge and orbital ordered state respectively. The right panel indicates the temperature dependent resistivity and magnetization for A-site disordered Sm$_{0.5}$Ba$_{0.5}$MnO$_{3}$ and Nd$_{0.5}$Ba$_{0.5}$MnO$_{3}$.}
\label{fig1}
\end{figure*}

	The half-doped manganites {\it Ln}$_{0.5}$Ba$_{0.5}$MnO$_{3}$ ({\it Ln} = Lanthanides) are one of the ideal systems to explore the physics of quenched disorder. The perovskite manganites in this chemical composition have two possible forms of the crystal structure depending on synthetic condition~\cite{ref9}. One is with the A-sites randomly occupied by {\it Ln}$^{3+}$ and Ba$^{2+}$ ions. The melt-quenched sample shows the complete solid solution of Ln and Ba ions on the A-sites with a simple cubic structure as the average structure. The other is the A-site ordered perovskite with the alternate stack of {\it Ln}O and BaO sheets along the c-axis with intervening MnO$_{2}$ sheets. This is due to the large difference in the ionic radii of {\it Ln}$^{3+}$ and Ba$^{2+}$. The respective MnO$_{2}$ sheets in this tetragonal form are free from random potential which would otherwise arise from the Coulomb potential or local strain from the random occupation of {\it Ln}/Ba on the A-sites.\par
The phase diagram~\cite{ref10,ref11} of ordered {\it Ln}$_{1/2}$Ba$_{1/2}$MnO$_{3}$ as a function of ionic radius is schematically shown by the dotted line in Fig. 1. The increase of ionic radius results in the reduction of Mn-O bond bending and hence the increases of the band width. With relatively small {\it Ln} ions from Dy to Sm, the CE-type charge ordered state with diagonal orbital stripes develops at low temperatures. With larger {\it Ln} ions from Nd to La, on the other hand, the ferromagnetic metal phase and, eventually, the A-type antiferromagnetic phase with ordering of $d_{x^{2}-y^{2}}$ orbitals are stabilized upon cooling the sample. These phases meet with each other at around {\it Ln} = Nd, forming a well defined multicritical point. When the solid-solution of {\it Ln}$^{3+}$ and Ba$^{2+}$ is formed, however, this multicritical region is drastically altered~\cite{ref11}. The solid line in Fig. 1 indicates that the phase diagram of A -site disordered {\it Ln}$_{0.5}$Ba$_{0.5}$MnO$_{3}$. It is clear that the phase transitions near the multicritical point are suppressed significantly. In contrast to the ordered {\it Ln}$_{0.5}$Ba$_{0.5}$MnO$_{3}$, the charge and orbital ordered phase goes away and alternatively a spin glass phase without long-range order in any charge and orbital sectors emerges. With increasing the ionic radius, the ferromagnetic metal phase takes over the spin glass phase around Nd. The A-type antiferromagnetic phase, however, does not show up at low temperatures. Besides, as compared with the A-site ordered systems, the ferromagnetic phase extends to the smaller ionic radius and the ferromagnetic transition temperature, {\it T}$_{C}$, is appreciably suppressed near the critical region.\par
The contrast between the A-site ordered and the disordered systems demonstrates a substantial influence of quenched disorder on the multiphase competition in the manganites. However, it should be emphasized here that the control parameter of the disordered system is not only the bandwidth but also the degree of disorder. By changing the {\it Ln} ion, the degree of disorder is inherently modified associated with the change in the ionic radius. It is difficult to identify which factor plays a dominant role in the unusual phase diagram of the disordered system. For example, the rapid decrease of {\it T}$_{C}$ in the ferromagnetic phase with approaching the critical region can be interpreted as an {\it enhanced} influence of quenched disorder in the critical region. Alternately, it may be understood in terms of the increased disorder with decreasing the ionic radius of {\it Ln} ions from those comparable with large Ba ions.\par
To separate the two contributions arising from the disorder and the bandwidth, we have employed hydrostatic pressure to modify the bandwidth alone, while keeping the degree of disorder constant. We have selected two A-site disordered compounds, Sm$_{0.5}$Ba$_{0.5}$MnO$_{3}$ and Nd$_{0.5}$Ba$_{0.5}$MnO$_{3}$, and applied hydrostatic pressures up to 10 GPa. We found a pressure induced spin glass insulator to ferromagnetic metal transition, which can be mapped onto the phase diagram obtained by examining the solid-solution effect. This clearly demonstrates that the unusual phase diagram of disordered {\it Ln}$_{0.5}$Ba$_{0.5}$MnO$_{3}$ arises from the bandwidth.\par
The A-site solid-solutions of {\it Ln}$_{0.5}$Ba$_{0.5}$MnO$_{3}$ ({\it Ln} = Sm and Nd) were grown by the floating-zone method in a single crystalline form. The starting polycrystalline rods were prepared by a standard solid-state reaction. The x-ray powder diffraction patterns of the resultant crystalline rods indicate the formation of single phase {\it Ln}$_{0.5}$Ba$_{0.5}$MnO$_{3}$ without any evidence for the A-site ordering. The resistivity and the magnetization data of disordered Sm$_{0.5}$Ba$_{0.5}$MnO$_{3}$ and Nd$_{0.5}$Ba$_{0.5}$MnO$_{3}$, at ambient pressure, are shown in the right panel of Fig. 1. The resistivity of disordered Sm$_{0.5}$Ba$_{0.5}$MnO$_{3}$ shows an insulating behavior down to the lowest temperature measured and no evidence for the long-range charge ordering is observed. Corresponding to the phase line {\it T}$_{SG}$ in Fig. 1, we observe a spin glass transition at 50 K in the magnetization but no appreciable anomaly at {\it T}$_{SG}$ is observed in the resistivity. In the Nd$_{0.5}$Ba$_{0.5}$MnO$_{3}$ sample, the resistivity shows a very rapid decrease below 140 K, which corresponds to the transition to a ferromagnetic metal phase. Note that the Curie temperature here is reduced to almost a half of the cation ordered Nd$_{1/2}$Ba$_{1/2}$MnO$_{3}$~\cite{ref10,ref11}. Hydrostatic pressures up to 10 GPa were applied to those two crystals by a cubic-anvil-type pressure apparatus~\cite{ref12}. The electrical resistivity and the AC susceptibility measurements were conducted within the pressure cell down to {\it T} = 4.2 K.\par

\begin{figure}[t]
\includegraphics[width=0.8\linewidth]{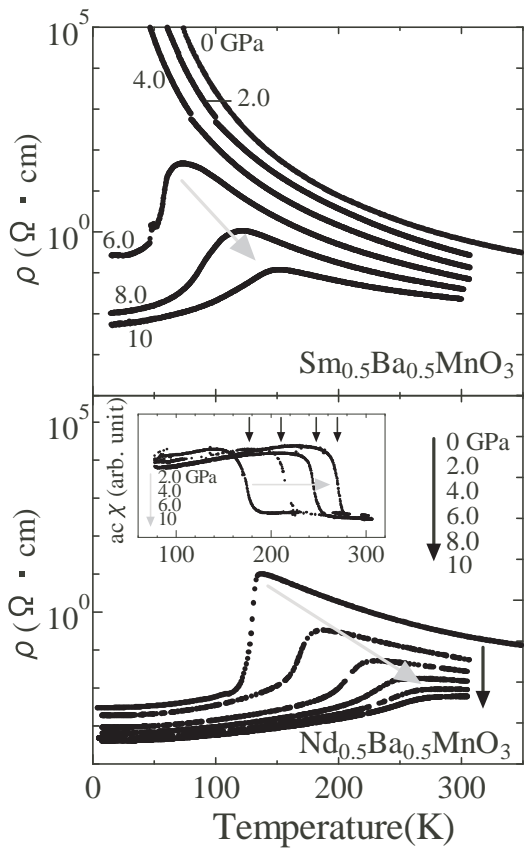}
\caption{Temperature dependent resistivities under various pressures for A-site disordered Sm$_{0.5}$Ba$_{0.5}$MnO$_{3}$ (top) and Nd$_{0.5}$Ba$_{0.5}$MnO$_{3}$ (bottom). The inset for Nd$_{0.5}$Ba$_{0.5}$MnO$_{3}$ indicates the AC susceptibility data for the same sample.}
\label{fig2}
\end{figure}

In the top panel of Fig. 2, the results of resistivity measurements on Sm$_{0.5}$Ba$_{0.5}$MnO$_{3}$ under various pressures were summarized. Up to 4 GPa, no appreciable change was observed except for the decrease of the magnitude of resistivity. The resistivity shows an insulating behavior down to 4.2 K, indicating that the system remains in the spin glass phase. Above 4 GPa, in contrast, a substantial change was observed at low temperatures. The resistivity shows a rapid decrease upon cooling, which is a characteristic behavior of the transition to a ferromagnetic metal. This clearly indicates that the system has experienced a pressure induced spin glass insulator to a ferromagnetic metal transition. Above the critical pressure, the transition temperature, {\it T}$_{C}$, goes up very rapidly with increasing pressure. These behaviors can be visually summarized as a phase diagram in Fig. 3. We immediately notice that the phase changes as a function of pressure are essentially the reproduction of those with the average ionic radius on the A-site.\par

\begin{figure}[b]
\includegraphics[width=0.8\linewidth]{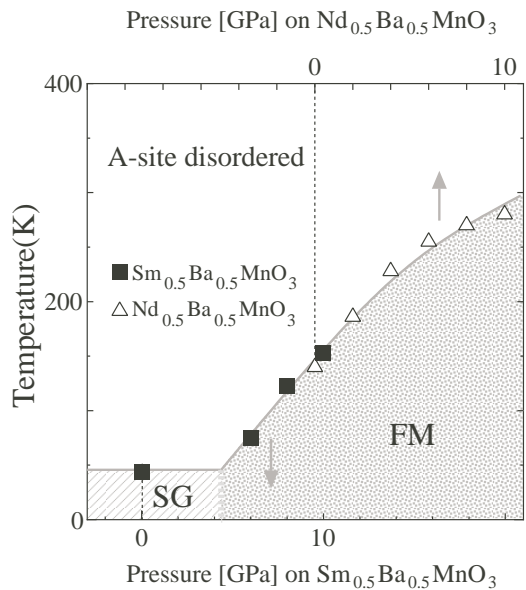}
\caption{Pressure phase diagram of A-site disordered Sm$_{0.5}$Ba$_{0.5}$MnO$_{3}$ and Nd$_{0.5}$Ba$_{0.5}$MnO$_{3}$, determined from the resistivity and the magnetic susceptibility data shown in Fig. 2. The pressure axis for Nd$_{0.5}$Ba$_{0.5}$MnO$_{3}$ is shifted by 10 GPa.}
\label{fig3}
\end{figure}

It is now clear that the dominant controlling factor of the unusual phase diagram in disordered {\it Ln}$_{0.5}$Ba$_{0.5}$MnO$_{3}$ is the bandwidth. Then, the pressure dependence of transition temperatures in disordered Nd$_{0.5}$Ba$_{0.5}$MnO$_{3}$ should be connected to that of Sm$_{0.5}$Ba$_{0.5}$MnO$_{3}$ by shifting the pressure axis. The bottom panel of Fig. 2 indicates the evolution of the temperature dependent resistivity and the AC susceptibility with increasing pressure for the A-site disordered Nd$_{0.5}$Ba$_{0.5}$MnO$_{3}$. {\it T}$_{C}$ defined as a temperature of resistivity anomaly shows a rapid increase with increasing pressure. At high pressures, the resistivity anomaly tends to become weaker, which make the determination of {\it T}$_{C}$ ambiguous. In the AC susceptibility, however, we are able to see the well defined anomalies associated with the ferromagnetic transition (indicated by arrows), which provides us with a solid estimate of {\it T}$_{C}$. Thus obtained pressure dependence of {\it T}$_{C}$ is plotted in Fig. 3. By shifting by 10 GPa, the {\it T}$_{C}$-{\it P} curve of disordered Nd$_{0.5}$Ba$_{0.5}$MnO$_{3}$ seems to be smoothly connected to that of Sm$_{0.5}$Ba$_{0.5}$MnO$_{3}$. This provides a further evidence for the dominance of pressure as the controlling parameter.\par
Another important point seen in Fig. 2 is that {\it T}$_{C}$ of Nd$_{0.5}$Ba$_{0.5}$MnO$_{3}$ tends to saturate with increasing pressure. In the saturation region, {\it T}$_{C}$ recovers to almost 300 K, which is close to those observed for ordered Ln$_{1/2}$Ba$_{1/2}$MnO$_{3}$ with {\it Ln} = Nd and Pr. These observations strongly suggest that the suppression of {\it T}$_{C}$ is much reduced once away from the critical region. Note again that the degree of disorder is kept constant in this experiment. We therefore conclude that the effect of quenched disorder is substantially {\it enhanced} when coupled with the (multi)criticality.\par
Recently, it was pointed out that the quenched disorder triggers a phase separation into ferromagnetic and charge ordered domains~\cite{ref13}. In their scenario, the formation of a clustered state with randomly oriented ferromagnetic clusters results in the decreases of the clean limit {\it T}$_{C}$ to a much lower {\it T}$_{C}$ in the ferromagnetic region. This accounts for the anomalous suppression of {\it T}$_{C}$ with approaching the critical region~\cite{ref14}. More recent works claim that the state below the clean limit {\it T}$_{C}$ is not statically clustered but homogeneous state with substantial critical charge/lattice fluctuations induced by quenched disorder~\cite{ref15}. Present pressure measurement alone cannot specify whether the phase separation is static or dynamic.\par
In summary, we have observed a pressure-induced switching of a spin glass into a ferromagnet in a half-doped manganite Sm$_{0.5}$Ba$_{0.5}$MnO$_{3}$ with quenched disorder, which is followed by a rapid increase of Curie temperature upon pressure. The remarkably rapid increase of Curie temperature was confirmed also in itinerant ferromagnet Nd$_{0.5}$Ba$_{0.5}$MnO$_{3}$ with a slightly larger ionic radius. Thus obtained pressure phase diagram can be mapped onto the unusual phase diagram of disordered {\it Ln}$_{0.5}$Ba$_{0.5}$MnO$_{3}$ as a function of ionic radius, indicating that the critical control parameter is the bandwidth under the presence of quenched disorder. It is now firmly established that the effect of quenched disorder is much more enhanced in the critical region, resulting in a stronger suppression of phase transition with approaching the multicritical point. The high pressure measurements have proved themselves as a remarkably useful probe to explore the physics of quenched disorder in multi-phase competing systems.

	The authors would like to thank Y. Motome, N. Nagaosa and Y. Tomioka for enlightening discussion. This work is partly supported by a Grant-in-Aid for Scientific Research from the Ministry of Education, Culture, Sports, Science and Technology, Japan and New Energy and Industrial Technology Development Organization (NEDO), Japan.\par


\end{document}